# GACELLE: GPU-accelerated tools for model parameter estimation and image reconstruction

**Authors**

[1,2]Kwok-Shing Chan, PhD, kchan2@mgh.harvard.edu

[1,2,3]Hansol Lee, PhD, HLEE89@mgh.harvard.edu

[1,2]Yixin Ma, PhD, YMA12@mgh.harvard.edu

[1,2]Berkin Bilgic, PhD, BBILGIC@mgh.harvard.edu

[1,2]Susie Y. Huang, MD, PhD, susie.huang@mgh.harvard.edu

[*1,2]Hong-Hsi Lee, MD, PhD, HLEE84@mgh.harvard.edu

[*4]José P. Marques, PhD, jose.marques@donders.ru.nl

**Affiliations**

[1]Department of Radiology, Athinoula A. Martinos Center for Biomedical Imaging, Massachusetts General Hospital, Charlestown, MA, United States

[2]Harvard Medical School, Boston, MA, United States

[3]Department of Biomedical Engineering, Ulsan National Institute of Science and Technology, Ulsan, South Korea

[4]Donders Institute for Brain, Cognition and Behaviour, Radboud University, Nijmegen, The Netherlands

[*]These authors contributed equally to this work.

**Correspondence to**

Kwok-Shing Chan, PhD

Athinoula A. Martinos Center for Biomedical Imaging, Charlestown, MA, United States

149 13th St, Charlestown, MA, USA

**Highlights**

- *GACELLE* is an open-source GPU framework for qMRI in MATLAB.
- Unified tool for fast estimation, regularisation, and uncertainty quantification.
- Up to 14,380-fold faster than CPU, enabling high-resolution multi-parametric qMRI.
- Improves precision, reproducibility, and noise suppression.
- Lowers barriers for qMRI adoption in research and clinical applications.

2Correspondence to Kwok-Shing Chan**Keywords**

GPU acceleration, Parameter estimation, Quantitative MRI, Markov chain Monte Carlo, Optimisation framework, Biophysical modelling

**Highlights**

- *GACELLE* is an open-source GPU framework for qMRI in MATLAB.
- Unified tool for fast estimation, regularisation, and uncertainty quantification.
- Up to 14,380-fold faster than CPU, enabling high-resolution multi-parametric qMRI.
- Improves precision, reproducibility, and noise suppression.
- Lowers barriers for qMRI adoption in research and clinical applications.

Correspondence to Kwok-Shing Chan

# Abstract

Quantitative MRI (qMRI) offers tissue-specific biomarkers that can be tracked over time or compared across populations; however, its adoption in clinical research is hindered by significant computational demands of parameter estimation. Images acquired at high spatial resolution and/or requiring fitting for multiple parameters often require lengthy processing time, constraining their use in routine pipelines and slowing methodological innovation and clinical translation.

We present *GACELLE*, an open source, GPU-accelerated framework for high-throughput qMRI analysis. *GACELLE* unifies a stochastic gradient descent optimiser (*askadam.m*) and a stochastic sampler (*mcmc.m*) under a common interface in MATLAB, enabling fast parameter mapping, improved estimation robustness via spatial regularisation, and uncertainty quantification. *GACELLE* prioritises accessibility and ease of integration: users only need to provide a forward signal model, while *GACELLE*'s backend manages computational parallelisation, automatic parameter updates, and memory-efficient batching. The stochastic solver performs fully vectorised Markov chain Monte Carlo with identical likelihoods on CPU and GPU, ensuring reproducibility across hardware.

Benchmarking demonstrates up to 451-fold acceleration for the stochastic gradient descent solver and 14,380-fold acceleration for stochastic sampling compared to CPU-based estimation, without compromising quantitative accuracy. We demonstrate *GACELLE*'s versatility on three representative qMRI models and on an image reconstruction task. Across these applications, *GACELLE* improves parameter precision, enhances test-retest reproducibility, and reduces noise in quantitative maps.

By combining speed, usability and flexibility, *GACELLE* provides a generalisable optimisation framework for medical image analysis. It lowers the computational barrier for advanced qMRI, paving the way for reproducible biomarker development, large-scale imaging studies, and clinical translation.

Correspondence to Kwok-Shing Chan

# 1. Introduction

Quantitative magnetic resonance imaging (qMRI) transforms MRI from a qualitative imaging tool into a quantitative probe of tissue microstructure, measuring physical parameters that reflect the local environment of (water) protons (Weiskopf et al., 2021). Established qMRI methods, such as relaxometry ($T_1$, $T_2$, $T_2^*$), quantitative susceptibility mapping (QSM) (QSM Consensus Organization Committee et al., 2024) and diffusion MRI (Le Bihan, 1995), are sensitive to iron content (Langkammer et al., 2010; Wood et al., 2005), myelination (Cho et al., 2022; Lutti et al., 2014), calcification (Chen et al., 2013; Deistung et al., 2013; Schweser et al., 2010) and tissue microstructure geometry (Basser et al., 1994; Beaulieu, 2002). The general improvement in data quality and acquisition efficiency has led to an increased diversity of contrast mechanisms that can be explored with MRI within a single scan session, stimulating the development of advanced multi-compartment models. Some examples of such models that characterise the MR signal across distinct tissue compartments or molecular environments to infer microstructural properties include: Standard Model of diffusion in white matter and its variants (Kaden et al., 2016; Novikov et al., 2019, 2018a, 2018b; Zhang et al., 2012); myelin water imaging (Chan and Marques, 2020; Deoni et al., 2008; MacKay et al., 1994; Nam et al., 2015; Oh et al., 2013; Prasloski et al., 2012); chemical exchange saturation transfer (CEST) for solute protons (van Zijl and Yadav, 2011; Ward et al., 2000), chemical shift-encoded imaging for lipid quantification (Dixon, 1984; Reeder et al., 2005). These measurements provide complementary information to morphological imaging and can reveal biological features not captured by conventional contrasts (Chan et al., 2025b; Schilling et al., 2022) at the expense of increasingly complex models.

Parameter mapping is central to qMRI, requiring the estimation of model parameters that link MR signals to underlying tissue properties. The most common approach is nonlinear least squares (NLLS) data fitting, which directly minimises the residuals between the measured and modelled signals. While widely used, NLLS is sensitive to initialisation (Nam et al., 2015), can converge to local minima, and becomes slow for large datasets or complex forward models. Markov chain Monte Carlo (MCMC) sampling (Alexander, 2008; Behrens et al., 2003; Hastings, 1970; Metropolis et al.,

Correspondence to Kwok-Shing Chan

1953) offers a supplementary alternative, providing posterior distributions for each parameter and thus richer uncertainty estimates, but at the cost of substantially longer runtimes. Both approaches are typically applied voxel-by-voxel sequentially, which, for high-resolution or multi-parameter models, can lead to runtimes extending to hundreds of hours on a standard CPU as the number of parameters to be estimated in a model grows (Chan and Marques, 2020). Parallel CPU computing can reduce runtimes, but improvements are constrained by hardware limits such as core count, memory bandwidth, and communication overhead. As models become more sophisticated to improve biological specificity, the number of parameters and the complexity of the forward model both increase, further slowing processing. Additionally, many qMRI models are numerically ill-conditioned (Chan and Marques, 2020; Jelescu et al., 2015; Lankford and Does, 2013; Pineda et al., 2005; West et al., 2019), with correlated parameters and multiple local minima that hinder convergence, particularly at low signal-to-noise ratios (SNR). Together, these factors make conventional CPU-based voxel-wise fitting a major bottleneck to the routine application of advanced qMRI models in large-scale studies and clinical settings.

Graphics Processing Unit (GPU) acceleration has emerged as a powerful strategy to address this computational bottleneck while maintaining estimation accuracy and precision. One major direction has been the implementation of conventional model fitting on GPUs using CUDA or Python (Chang et al., 2014) . Harms et al. implemented diffusion microstructure models, including NODDI (Zhang et al., 2012) and CHARMED (Assaf et al., 2004) on GPU (Harms et al., 2017), showing the gradient-free Powell conjugate-direction optimisation provided improved runtime, fit accuracy, and precision, with a follow-up work extending the GPU acceleration to MCMC sampling approaches (Harms and Roebroeck, 2018). Hernandez-Fernandez et al. extended GPU acceleration from microstructure estimation to tractography and connectome reconstruction, enabling whole-brain datasets to be processed in seconds to minutes while maintaining quantitative agreement with CPU solutions (Hernandez-Fernandez et al., 2019). These implementations demonstrate the efficiency of GPU processing but are often tied to fixed model implementations and require specialised programming, which limits adoption, extension, and development of the model space.

Correspondence to Kwok-Shing Chan

A second direction is the use of deep learning methods to bypass iterative fitting altogether. Artificial neural networks (ANNs) have been applied to qMRI parameter estimation tasks, including myelin water imaging (Jung et al., 2021; Lee et al., 2019), diffusion-relaxation model of white matter (de Almeida Martins et al., 2021), intravoxel incoherent motion model (Barbieri et al., 2020), and permeability for white matter microstructure imaging (Nedjati-Gilani et al., 2017), providing rapid inference after offline training. These supervised methods significantly reduce computation time and often improve robustness to noise, especially in low SNR regimes. More recently, physics-informed self-supervised strategies, such as the diffusion Model OptimizatioN with deep learning (DIMOND) (Li et al., 2024) and implicit neural representations (INRs) (Hendriks et al., 2025), have enabled fast estimation without the need for explicit ground-truth parameter maps. Hybrid approaches such as $\mu$GUIDE (Jallais and Palombo, 2024) further integrate deep learning with uncertainty modelling to improve reliability across acquisition conditions. Although learning-based methods offer significant acceleration, supervised approaches can be biased by the availability of the training data (Gyori et al., 2022), and often require re-training or adaptation for each acquisition protocol, scanner vendor, or clinical population, which can limit their direct applicability in heterogeneous settings. In contrast, self-supervised methods that optimise directly on the target data without offline training may be more robust to such variations, improving generalisability across diverse imaging scenarios.

To provide a GPU-accelerated option that balances speed, usability, and flexibility, we developed *GACELLE*, a MATLAB-based framework for high-throughput qMRI parameter estimation. In contrast to most GPU-accelerated tools, which are typically implemented in CUDA or Python, *GACELLE* operates in MATLAB, lowering the barrier for researchers who are already working in this environment but may not have experience with GPU programming. The framework integrates two solvers: (1) a stochastic gradient descent optimiser (*askadam.m*) that fits all voxels simultaneously within a single objective function, and (2) a stochastic MCMC sampler (*mcmc.m*). Both solvers support user-defined forward models and loss functions, allowing rapid adaptation to new acquisition protocols or models. By combining ease of use, flexibility, and GPU-level performance, *GACELLE* provides a practical solution for accelerating model-based qMRI analysis in research and potentially clinical workflows.

Correspondence to Kwok-Shing Chan

# 2. Methods

## 2.1. Dependency, installation and documentation

*GACELLE* is a MATLAB-based framework for high-throughput qMRI parameter estimation, supporting both stochastic gradient descent optimisation (objective function minimisation) and stochastic inference via MCMC sampling. The framework is designed to leverage GPU to substantially accelerate computation while minimising the need for user-side GPU programming.

The current implementation (v1.0) supports MATLAB R2022b and later versions on Linux and Windows operating systems. GPU acceleration requires compatible hardware and drivers, as specified by the MATLAB GPU Computing Requirements (https://www.mathworks.com/help/parallel-computing/gpu-computing-requirements.html).

*GACELLE* is distributed under the GNU General Public License v3.0 (GPL-3.0) and is available as open source on GitHub (https://github.com/kschan0214/gacelle). Current supported signal models include: $T_2^*$-based (Nam et al., 2015) and multi-compartment relaxometry based (Chan and Marques, 2020) myelin water imaging, axon diameter mapping using spherical mean technique (AxCaliberSMT) (Assaf et al., 2008; Fan et al., 2020), multi-compartment microscopic diffusion imaging (Kaden et al., 2016), soma and neurite density imaging (SANDI) (Palombo et al., 2020), and neurite exchange imaging (NEXI) (Chan et al., 2025a; Jelescu et al., 2022; Uhl et al., 2024). Comprehensive documentation, including installation instructions, API descriptions and tutorials, is provided online (https://gacelle.readthedocs.io/en/latest/index.html).

## 2.2 Parameter estimation solvers

*GACELLE* includes two solvers for high-throughput parameter estimation:
(1) An adaptive gradient moment optimiser that minimises a global objective function (*askadam.m*), and

(2) A stochastic inference method based on MCMC sampling (Hastings, 1970; Metropolis et al., 1953) (*mcmc.m*).

Both solvers are implemented to process multiple voxels/samples in parallel on a GPU.

## 2.2.1. Stochastic gradient descent optimiser: *askadam.m*

Conventional NLLS fitting estimates a qMRI model parameter set $\theta$ by minimising the difference between the measured signal $s_{meas,r}$ and the model prediction $s_r$ for each voxel $r$ independently:

$$\underset{\theta}{\mathrm{argmin}} \left\| w[s_{meas,r}(\phi) - s_r(\phi, \theta)] \right\|_2^2 \quad [Eq.\,1]$$

where $w$ is an optional weighting term, typically reflecting voxel SNR, and $\phi$ is the acquisition parameters. While effective, this voxel-wise approach must be repeated sequentially across the entire imaging volume.

In *GACELLE*, *askadam.m* reformulates this as a single optimisation problem over **the entire imaging volume**:

$$\underset{\theta}{\mathrm{argmin}} \left\| MW[S_{meas}(\phi) - S(\phi, \theta)] \right\|_{L1|L2} \quad [Eq.\,2]$$

where $S_{meas}(\phi)$ and $S(\phi, \theta)$ are the (multi-dimensional) measurement data and model prediction for all input voxels, respectively, $W$ is the volumetric weights, and $M$ is the signal mask. This formulation aggregates all voxels within $M$ into a single loss function, enabling simultaneous gradient computation for all parameters using MATLAB's automatic differentiation, similar to neural network training. This operation greatly simplifies how a new signal model can be incorporated into the framework, as the Jacobian matrix at the solution (i.e., the partial derivatives of the objective function with respect to $\theta$) no longer needs to be specified explicitly, which is not always trivial or analytically exists depending on the complexity of the forward model. The shared forward model across voxels also allows vectorised GPU evaluation, with both gradient computation and parameter updates executed on the GPU (Figure 1). Such whole volume minimisation problems are common in MRI in the context of linear problems, such as conjugate gradient (CG)-SENSE image reconstruction (Pruessmann et al., 2001), model-based image reconstruction (Fessler, 2010), sub-space image reconstruction (Dong et al., 2020; Wang et al., 2022), QSM (QSM Consensus Organization Committee et al., 2024) and microstructure parameter estimation (Daducci et al., 2015) .

The solver supports multiple gradient descent optimisers for parameter update, including Adaptive Moment Estimation (Adam) (Kingma and Ba, 2014), Stochastic Gradient Descent with Momentum (SGDM) and Root Mean Square Propagation (RMSProp). Users can control the step size of the parameter update using the *initialLearnRate* option. For each optimiser, two loss functions are available, including L1-norm and L2-norm. The optimisation iterations can be terminated when one of the following stopping criteria is fulfilled:

1. The loss value is smaller than the maximum threshold specified in the *tol* option;
2. The maximum number of optimisation iterations specified in the *iteration* option is reached; and
3. The gradient of the loss as a function of the iterations is smaller than the value specified in the *convergenceValue* option. This gradient is defined as the slope of the loss value over a window of the last N iterations, where N>1. Users can adjust the window size N using the *convergenceWindow* option. The default value of *convergenceWindow* is 20 iterations to be robust to the small fluctuation in the loss value changes during the optimisation process.

The default settings are designed for robustness but can be tuned for specific applications. The full list that controls the optimisation setting is available on the Documentation website.

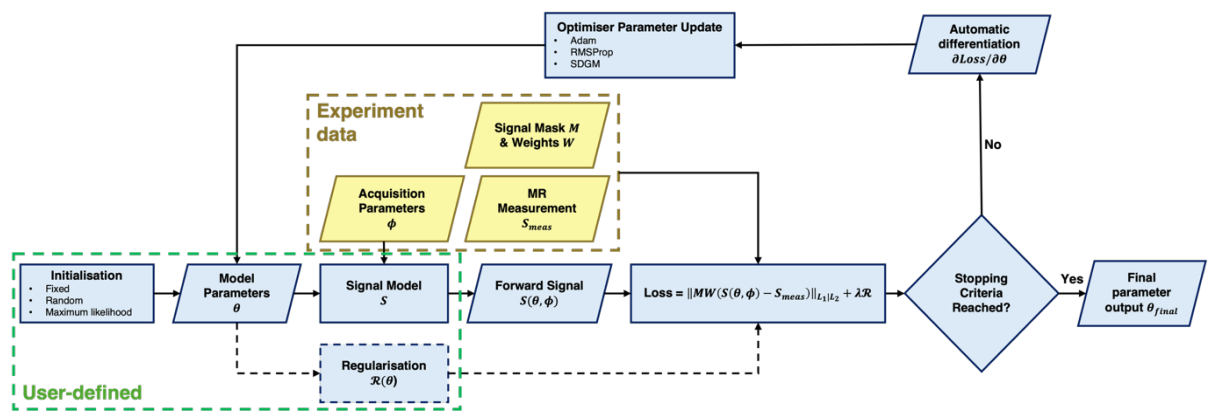

Figure 1: An illustration of the *askadam.m* solver optimisation process. Green frame highlights the processing components that users can adjust. Yellow frame indicates the input data.

Correspondence to Kwok-Shing Chan

## 2.2.2. Stochastic inference solver: *mcmc.m*

MCMC is a class of numerical algorithms designed to approximate the probability distribution of a target random variable. These methods generate a sequence of random variables forming a Markov chain, where each state depends only on the preceding one, and employ random sampling to estimate the underlying distribution. MCMC can be used in Bayesian inference for model parameter estimation from the qMRI context. For example, one can infer the expected value and estimation uncertainty from the posterior distribution of model parameters given the measurement data (Harms and Roebroeck, 2018).

There are many MCMC algorithms available, and *GACELLE* includes two MCMC algorithms: (1) Metropolis-Hastings (Hastings, 1970; Metropolis et al., 1953), and (2) affine-invariant ensemble sampling (Goodman and Weare, 2010), that are customised for high-throughput estimation tasks (*mcmc.m*). For both, the likelihood is Gaussian, and the proposed distribution is normal. The implementation is fully vectorised, with all chains and likelihood computations processed in parallel. This design leverages MATLAB's optimised matrix operations and GPU acceleration, avoiding loop-based evaluation across voxels.

When using the Metropolis-Hastings algorithm, users can adjust the individual step size of the proposal for the model parameters using the *xStepSize* option. In contrast, a single value *stepSize* is used to control the step size of all parameters in the affine-invariant ensemble method. Other options that control the algorithm settings include:

- *iteration*: number of iterations for the MCMC sampling process;
- *burnin*: the ratio of iterations that will be discarded at the beginning of the process to mitigate the effect of the choice of the starting position;
- *thinning*: the sampling interval of the chain to mitigate the effect of autocorrelation in the posterior distribution samples;
- *Nwalkers*: number of random walkers in the ensemble (for the affine-invariant ensemble method only).

In contrast to *askadam.m*, the GPU implementation of MCMC is numerically identical to the CPU version, ensuring reproducibility across hardware configurations.

## 2.3. Integration of a new signal model

### 2.3.1 Forward Function Design

*GACELLE* is designed to minimise the technical overhead of incorporating new qMRI signal models. Users only need to define the forward signal model as a MATLAB function handle, which can be used directly by either solver (*askadam.m* or *mcmc.m*; Figure 2).

As an example, consider a single-compartment $R_2^*$ relaxometry modelled as:
$$S(t, S_0, R_2^*) = S_0 e^{-tR_2^*} \; [Eq. 3]$$
where $t$ is the echo time, $S_0$ is the $T_1$-weighted signal and $R_2^*$ is the effective transverse relaxation rate. Integration into *GACELLE* involves:

**Step 1**: Define model parameter names (e.g., $S_0$ and $R_2^*$) as fields of the function's input structure (*parameter.S0* and *parameters.R2star),* as illustrated in Figure 2.

**Step 2**: Implementing the forward model equation, with acquisition parameters (i.e., echo times, *t*) passed as additional inputs to the function. These acquisition parameters can be organised into structured arrays of arbitrary dimensions, ideally formatted for direct use in matrix operations alongside the parameters to be estimated.

**Step 3**: Setting solver-specific optimisation options, such as parameter bounds, initialisation, loss function, stopping criteria, etc.

**Step 4**: Calling the solver function (*askadam.m* or *mcmc.m*) for parameter estimation. The same forward function can be applied to both solvers without modification.

The output formats of *askdam.m* and *mcmc.m* differ, reflecting the point-estimate nature of *askadam.m* (in the example of Figure 2, the final results from *askadam.m* are organised as *out.final.S0* and *out.final.R2star*), while *mcmc.m* additionally characterises the full fitting distribution (e.g., mean and standard deviation of the posterior distributions are stored as *out.mean.R2star* and *out.std.R2star*). Detailed information about the output organisation is available in the API section in the online documentation.

Figure 2: Example workflow for parameter estimation in *GACELLE*. (a) Script for fitting using the stochastic gradient descent solver (*askadam.m*). (b) Forward model implementation for single-compartment $R_2^*$ relaxometry (Eq. [3]). (c) Script for fitting using the stochastic inference solver (*mcmc.m*). The same forward function is compatible with both solvers, highlighting the minimal adaptation required when switching estimation strategies.

## 2.3.2 Data handling

Quantitative MRI data are typically multi-dimensional, with the first 3 dimensions representing spatial coordinates and additional dimensions corresponding to acquisition parameters (e.g., echo time, flip angle or *b*-value). In *GACELLE*, model parameters are passed to the forward function in the native N-dimensional shape. For example, for a 4D multi-echo gradient-echo (GRE) dataset (x, y, z, t) in $R_2^*$ mapping, parameters $S_0$ and $R_2^*$ are provided as 3D matrices. The forward function must return a signal array matching the dimensions of the measured data (i.e., 4D in this case). Signal masking is handled internally during loss computation (see Figure 3a).

GPU memory constraints can arise with large multi-dimensional datasets. *GACELLE* addresses this by allowing optional masking of model parameters *before* forward model evaluation (Figure 3b). When the *isOptimiseMemory* option is enabled, computations are restricted to voxels within the mask, reducing memory usage.

Parameters can be reshaped back to their original size if spatial regularisation is applied (see also Section 2.4). Alternatively, users can also apply masking externally and provide only the masked data for the use of *GACELLE* in estimation.

For *askadam.m*, forward functions must be compatible with MATLAB *dlarray* objects to enable automatic differentiation. The MATLAB documentation website (https://www.mathworks.com/help/) provides a list of supported operations for *dlarray* objects. For *mcmc.m*, which does not require differentiation, standard GPU array operations (*gpuArray*) are supported.

## 2.4. Incorporating regularisation with *askadam.m*

A key advantage of the stochastic gradient descent solver (*askadam.m*) is that the loss function can be extended to include a global regularisation term $R$ during optimisation. This allows users to promote desired parameter map properties (e.g., sparsity or spatial smoothness) and improve convergence, an approach widely used in imaging reconstruction (Knoll et al., 2011), QSM (QSM Consensus Organization Committee et al., 2024) and multi-compartment parameter fitting (Orton et al., 2014; Pasternak et al., 2008). The objective function in Eq. [2] can be generalised into:

$$\underset{\theta}{\operatorname{argmin}} \| Mw[S_{meas}(\phi) - S(\phi, \theta)] \|_{L1|L2} + \lambda R \; [Eq. 4]$$

where $\lambda$ is a user-specified regularisation weight. Because *askadam.m* updates all voxel parameters simultaneously, the regulariser is updated at each iteration, ensuring consistency throughout the optimisation.

Currently, *GACELLE* provides built-in 2D and 3D total variation (TV) regularisation, which can be applied to one or more parameter maps:

$$R = |\nabla \theta| \; [Eq. 5]$$

Users may also implement a custom regularisation function (e.g., L1-based or L2-based), provided they return a scalar loss value. This flexibility allows integration of prior knowledge, such as parameter distribution from literature, or compound regulariser (e.g., edge-preserved TV combined with tissue-specific regularisation like in (Liu et al., 2018)).

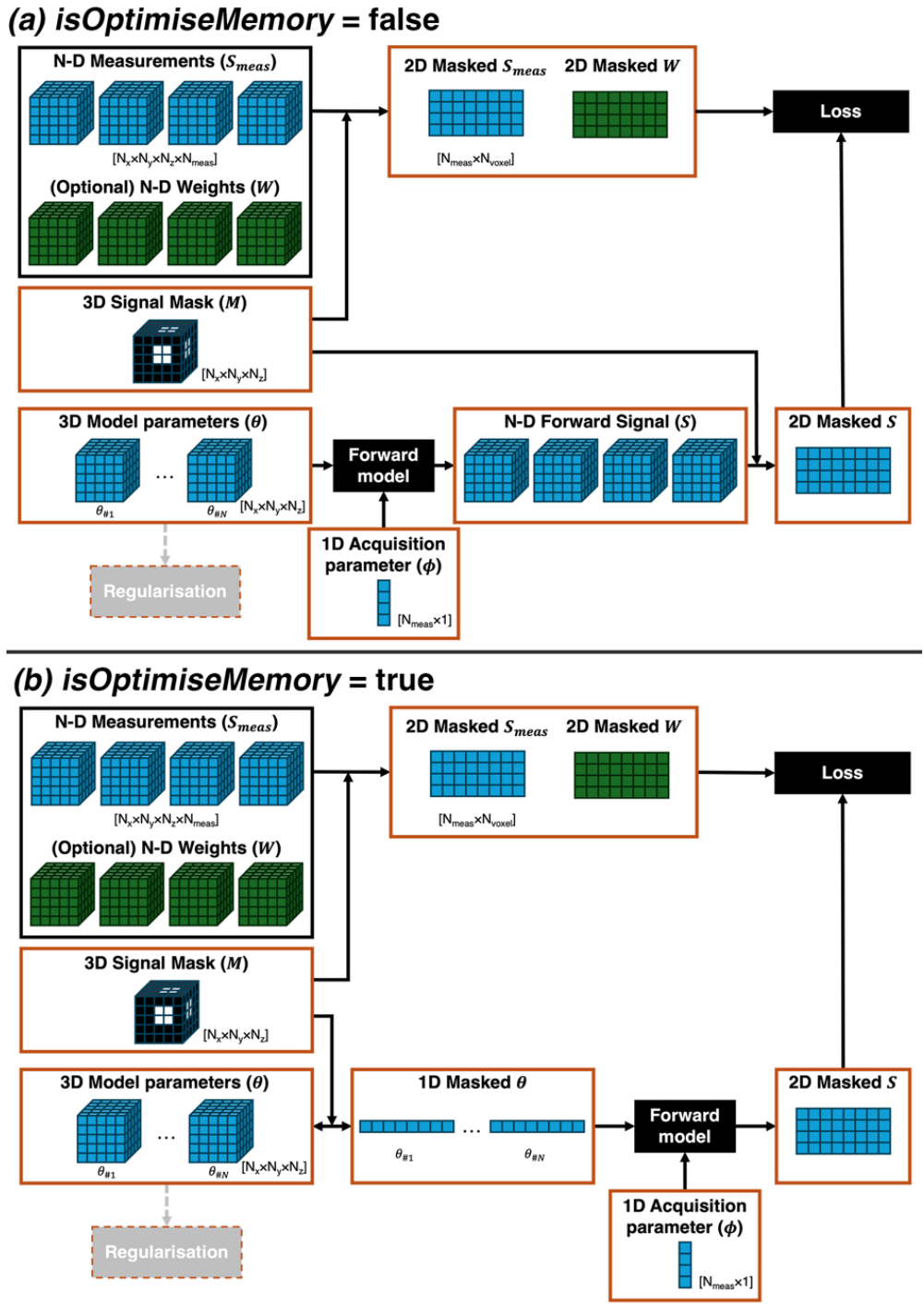

Figure 3: Data handling options in *GACELLE*. When *isOptimiseMemory* = false, the forward model operates directly on the input data in its original shape, suitable for tasks such as image reconstruction where masking is not essential. When *isOptimiseMemory* = true, masking is applied to model parameters before forward model evaluation, reducing GPU memory demands by restricting computations to voxels within the mask. Orange boxes indicate data resident on the GPU during optimisation.

## 2.5. Speed test

The computational performance of *GACELLE* is influenced by model complexity, implementation strategy, data size and hardware specifications. To benchmark its speed, we compared the *GACELLE* solvers against conventional voxel-wise NLLS fitting and MCMC sampling. As test cases, we used the AxCaliber model with the spherical mean techniques (AxCaliberSMT) (Fan et al., 2020) and the NEXI model (Jelescu et al., 2022), which both estimate a moderate number of tissue parameters (i.e., 4 in total). For AxCaliberSMT, the fitting parameters include axon diameter, neurite signal fraction $f_{neurite}$, free water signal fraction $f_{CSF}$ and the radial diffusivity of the hindered extra-cellular water ($D_{e,R}$). The intrinsic diffusivity of restricted axonal water, as well as the longitudinal diffusivities of axonal and extra-cellular water, were fixed to 1.7 $\mu$m$^2$/ms, and the diffusivity of free water was fixed to 3 $\mu$m$^2$/ms, following (Fan et al., 2020). For NEXI, the fitting parameters include neurite signal fraction $f_n$ and diffusivity $D_n$, extracellular diffusivity $D_e$ and the exchange rate from neurite to extracellular space $r_n$. Our implementation projects the NEXI signal kernel onto Legendre polynomials (Novikov et al., 2018b), enabling the use of higher-order rotationally invariant spherical harmonic components in data fitting (Chan et al., 2025a).

Synthetic diffusion signals were generated using the acquisition protocol described in Supplementary Table S1 for Demo #1 (NEXI; 15 b-values; see also Section 2.6.1) and Demo #2 (AxCaliberSMT; 16 b-values; see also Section 2.6.2) for 2.5×10$^5$ samples, with SNR of 50 at b=0 using the NEXI forward model, and SNR of 30 at b=0 using AxCaliberSMT. Unlike voxel-wise CPU estimation, where runtime scales linearly with data size, GPU runtime scaling is nonlinear. To characterise this, we performed parameter estimation with varying sample counts [10$^2$, 10$^3$, 10$^4$, 10$^5$, 2.5×10$^5$] in *GACELLE*. For *askadam.m*, the maximum iteration number was set to 10000, and the convergence value was set to 10$^{-8}$. For *mcmc.m*, we used the same optimisation setting as in Table 1 for both AxCaliberSMT and NEXI. Each condition was repeated 10 times, and the mean processing time was used to compute the average time per sample. For CPU-based NLLS and MCMC sampling, only 10$^3$ samples were processed, as runtime scales approximately linearly with the number of samples. To enable fair speed comparison, we used a single CPU core, allowing theoretical extrapolation to multi-threaded performance (e.g., using MATLAB's *parfor*; assuming negligible communication overhead). All computations were performed on an AMD EPYC 7313 16-core processor equipped with an NVIDIA A40 GPU.

## 2.6. Example usages of *GACELLE* in qMRI

To illustrate the flexibility of *GACELLE* across diverse qMRI problems, we present four representative use cases. Common statistical analyses included:

- Agreement between solvers assessed by Bland–Altman analysis.
- Precision evaluated by interquartile range (IQR) and coefficient of variation (CoV) across ROIs.
- Test–retest reproducibility assessed by Pearson correlation when applicable.

### 2.6.1. Demo #1: *askadam.m* with regularisation - Neurite Exchange Imaging (NEXI)

This example demonstrated the flexibility of *askadam.m* in applying spatial regularisation for microstructure model fitting, including both volumetric and surface-based parameter estimation, and incorporating prior-based regularisation to improve robustness at lower SNR.

NEXI model fitting was performed using dMRI data previously acquired for (Chan et al., 2025a) on the 3T Connectome 2.0 scanner (MAGNETOM Connectom.X, Siemens Healthieers, Erlangen, Germany), equipped with the maximum gradient strength of 500 mT/m and maximum slew rate of 600 T/m/s (Huang et al., 2021; Ramos-Llordén et al., 2025), using an in-house built 72-channel head coil for signal reception (Mahmutovic et al., 2025). Diffusion protocol included: 2D spin-echo echo-planar-imaging (EPI) with three diffusion times and 15 *b*-values, either 64 (high-SNR dataset; 10 healthy participants) or 32 (low-SNR dataset; 5 healthy participants) gradient directions per shell, from which the intercompartmental exchange time between neurite and extracellular space was estimated. $T_1$-weighted anatomical images were also acquired for cortical parcellation. Full acquisition, participant demographics, and preprocessing details were reported in (Chan et al., 2025a).

**Volumetric parameter estimation**

Rotationally invariant dMRI signal (zeroth and second order) (Novikov et al., 2018b) was fitted to the corresponding NEXI model using:

a) voxelwise NLLS on CPU (NEXI$_{NLLS}$)
b) *askadam.m* without regularisation (NEXI$_{aA}$)
c) *askadam.m* with 2D or 3D TV regularisation on the neurite signal fraction map $f_n$ (NEXI$_{aA2DTV}$, NEXI$_{aA3DTV(fn)}$)
d) *askadam.m* with 3D TV applied jointly to $f_n$ and exchange rate from neurite to extracellular water $r_n$ (NEXI$_{aA3DTV(fn,rn)}$)

Grey matter ROIs were extracted using SynthSeg (Billot et al., 2023) on $T_1$-weighted data, then registered to the individual diffusion space. For each ROI, median parameter values were computed to minimise any registration inaccuracy. Agreement between solvers was assessed using Bland-Altman analysis; precision was quantified by IQR and group-level coefficient of variation (CoV) $CoV = \sigma(\theta)/\bar{\theta}$.

**Surface-based parameter estimation**

Spatial regularisation assumes that spatially adjacent voxels share similar microstructural properties. In highly folded structures such as the cortex, however, nearby voxels in (volumetric) Euclidean space can belong to different cortical layers or gyri/sulci, leading to potential mixing of unrelated tissue properties, particularly at low spatial resolution.

To address this, we demonstrated here a surface-based fitting approach using *askadam.m*. Rotationally invariant dMRI signals from the low SNR dataset were first projected from volumetric space to the *fsaverage* surface at the mid-cortical depth, generated using FreeSurfer's *recon-all* pipeline (Fischl, 2012). Each surface mesh vertex represents a spatially coherent location in cortical space, where neighbourhood relationships are defined by the mesh connectivity rather than Euclidean distance in volume space.

Total variation regularisation was applied along vertex neighbourhoods (NEXI$_{aASurfaceTV}$). For each vertex $v_i$, the surface TV term was computed as:

$$R(\theta) = \sum_{v_j \in \mathcal{N}(v_i)} |\theta(v_i) - \theta(v_j)| \quad Eq.\,[5]$$

where $\mathcal{N}(v_i)$ is the set of vertices directly connected to $v_i$ by an edge in the surface mesh. This definition ensures that regularisation is applied across geodetically adjacent vertices, avoiding spurious smoothing across unrelated tissue. The regularisation weight ($\lambda$) was empirically set to $10^{-5}$ for $f_n$ based on preliminary experiments optimising noise suppression without over-smoothing.

As an additional demonstration of the flexibility of the regularisation framework, we applied a simple prior-based regularisation term to the low-SNR dataset using parameter distributions obtained from the high-SNR dataset. This illustrates that *askadam.m* can accommodate arbitrary custom regularisation functions in addition to built-in TV. The prior-based term was defined as:

$$R(\theta) = \left| \frac{\theta - \mu_{ROI}}{\sigma_{ROI}} \right| \quad [Eq.\,6]$$

where $\mu_{ROI}$ and $\sigma_{ROI}$ were the group mean and standard deviation of the corresponding model parameter of an ROI, and were derived from the volumetric fitting of NEXI$_{aA}$ to a high-SNR dataset. This example is not intended as a comprehensive prior optimisation study, but to show that the framework can integrate prior distributions from external datasets to stabilise estimation when SNR is reduced. All optimisation settings with various solvers are summarised in Table 1.

Correspondence to Kwok-Shing Chan

Table 1: Summary of optimisation parameters and solvers used for Demo#1-4. All the regularisation parameters were derived empirically.

| Demo | #1<br>NEXI | #2<br>AxCaliberSMT | #3<br>MCR-MWI | #4<br>Iterative SENSE recon |
|---|---|---|---|---|
| **Solver** | **Volumetric fitting**<br>(1) NLLS<br>(2) *askadam.m*<br>(3) *askadam.m* + $\lambda\|\nabla_{2D}f_n\|$<br>(4) *askadam.m* + $\lambda\|\nabla_{3D}f_n\|$<br>(5) *askadam.m* + $\lambda\|\nabla_{3D}f_n\|$ + $\lambda\|\nabla_{3D}r_n\|$<br><br>**Vertex fitting**<br>(1) NLLS<br>(2) *askadam.m*<br>(3) *askadam.m* + $\lambda\|\nabla_{\text{surface}}f_n\|$<br>(4) *askadam.m* + $\sum_\theta \lambda_{Prior}\left\|\frac{\theta-\mu_{ROI}}{\sigma_{ROI}}\right\|$ | (1) MH (CPU)<br>(2) MH (GPU)<br>(3) Ensemble (GPU) | (1) NLLS<br>(2) *askadam.m* | (1) LSQR<br>(2) LSQR + Tikhonov<br>(3) *askadam.m* with L1-norm<br>(4) *askadam.m* with L2-norm<br>(5) *askadam.m* with L1-norm + $\lambda\|\nabla_{2D}I\|$ |
| **Optimisation setting** | - 4000 max. iterations<br>- L1 loss<br>- $10^{-8}$ convergence<br><br>**Volumetric fitting**<br>- $\lambda$ = 0.001<br><br>**Vertex fitting**<br>- $\lambda$ = $10^{-5}$<br>- $\lambda_{Prior}$ = $10^{-4}$ | - 10% burn-in<br>- Thinning = 100 iterations<br><br>**MH (900 posterior samples)**<br>- 4 independent proposals<br>- 25000 iterations per proposal<br><br>**Ensemble (900 posterior samples)**<br>- 50 walkers<br>- 2000 iterations | - $10^4$ max. iterations<br>- L1 loss<br>- $10^{-8}$ convergence | - 500 max. iterations<br>- $10^{-8}$ (L1) or $10^{-9}$ (L2) convergence<br>- $\lambda$ = $10^{-4}$ (Tikhonov) / 0.002 (*askadam.m*) |

## 2.6.2. Demo #2: using *mcmc.m* for AxCaliber with spherical mean technique (AxCaliberSMT)

This example demonstrated the application of the stochastic MCMC solver (*mcmc.m*) to microstructure parameter estimation in a complex model of axon diameter mapping using AxCaliberSMT (Fan et al., 2020), comparing CPU- and GPU-based Metropolis–Hastings's implementations as well as the affine-invariant ensemble samplers.

Diffusion MRI data were acquired on the 3T Connectome 2.0 scanner using the 72-channel head coil (Mahmutovic et al., 2025). Imaging experiments were performed on 10 healthy volunteers (26-49 years; 9 females, 1 male). The study was approved by the local ethics committee, and written informed consent was obtained from all participants. The imaging protocol comprises:

- Whole-brain $T_1$-weighted scan using MPRAGE with 0.9 mm isotropic resolution, acquisition time = 5.5 min;
- 2D spin-echo EPI-dMRI, multi-band factor of 2, GRAPPA factor of 2, partial Fourier of 6/8, resolution = 2 mm isotropic, TR/TE = 3600/54 ms using the diffusion scheme laid out in Supplementary Table S1. Non-diffusion-weighted images (b=0) were acquired interspersed every 16 diffusion-weighted images (DWIs). Total acquisition time = 56 min.

Data were processed using the same pipeline as in Section 2.6.1. AxCaliberSMT estimation was performed using:

(1) Metropolis-Hastings algorithm on CPU ($MH_{CPU}$)
(2) Metropolis-Hastings algorithm on GPU via *GACELLE* ($MH_{GPU}$)
(3) Affine-invariant ensemble samplers on GPU via *GACELLE* (Ensemble).

Details of the optimisation setting can be found in Table 1. For comparison, $MH_{CPU}$ was repeated with an independent proposal scheme on one subject, yielding a second MCMC posterior distribution. We compared the results from these two proposal schemes and used this comparison as the baseline for evaluating the CPU and GPU implementation. Posterior means and standard deviations were computed for each parameter, representing point estimates and uncertainties. White matter ROIs were segmented using TractSeg (Wasserthal et al., 2018). Agreement among implementations ($MH_{CPU}$, $MH_{GPU}$, Ensemble) was evaluated using Bland–Altman analysis.

## 2.6.3. Demo #3: using *askadam.m* with intractable model: multi-compartment relaxometry for myelin water imaging (MCR-MWI)

This example illustrates the ability of *askadam.m* to fit computationally intensive models that lack a closed-form solution by integrating a neural network surrogate for forward signal simulation. We chose the MCR-MWI model (Chan and Marques, 2020), which estimates myelin water fraction (MWF) by modelling relaxation and susceptibility differences between myelin water and intra-/extra-cellular water using multi-echo, variable flip angle GRE imaging. This model incorporates an extended phase graph with exchange (EPG-X) (Malik et al., 2017) to account for magnetisation exchange effect and hardware imperfections such as gradient and RF spoiling on the measured GRE signal. Utilising EPG-X in data fitting is computationally expensive because the EPG-X simulation must be repeated multiple times to simulate the RF effects on the steady-state signal at each optimisation iteration.

To address this computational bottleneck, we trained a multi-layer perceptron (MLP) neural network to approximate the EPG-X simulated $T_1$-weighted signal. The ANN surrogate was then combined with analytical $T_2^*$ decay and magnetic susceptibility frequency shift terms to form the complete forward model used in *askadam.m*. This approach allows efficient GPU-accelerated optimisation while preserving the model's physical realism. Details of the ANN architecture, training procedure, and validation are provided in the Supplementary Materials A.

The demonstration used publicly available test-retest GRE data (3 subjects, two sessions; https://doi.org/10.34973/3qcy-jt86), acquired at 3T with multi-echo, variable flip angle GRE sequences (see (Chan et al., 2023) and Supplementary Table S1 for full parameters). Briefly, data were acquired with three different GRE protocols (varying TEs and TR) so that various scanner setups were evaluated. Parameter estimation was performed using both CPU-based NLLS and GPU-based *askadam.m*. White matter ROIs were extracted using TractSeg (Wasserthal et al., 2018), and agreement between methods was assessed using the Bland-Altman analysis. Test-retest reproducibility was evaluated using Pearson correlation, consistent with prior analysis (Chan et al., 2023). We used the MWF maps derived from voxelwise NLLS fitting of MCR-MWI incorporating diffusion prior (i.e., MCR-DIMWI; (Chan and Marques, 2020)) as our visual references, which were shown to have higher SNR estimation in white matter (Chan et al., 2023).

### 2.6.4. Usage #4: using *askadam.m* for image reconstruction

Although *GACELLE* is primarily designed for high-throughput qMRI parameter estimation, *askadam.m* can also be applied to other large-scale optimisation problems in MRI, such as image reconstruction with parallel imaging acquisition. To illustrate this versatility, we used *askadam.m* to reconstruct highly accelerated multi-echo GRE acquisition with controlled aliasing patterns shifted across the echo time dimension (Wang et al., 2023). Data were acquired at 3T with the following parameters: TR = 35 ms, $TE_1/\Delta TE/TE_6$ = 3/5/28 ms, flip angle = 15°, acceleration $R_y$ = 1, $R_z$ = 9, CAIPI z-shift = 3 and CAIPI z-TE-shift = 2, 1 mm isotropic resolution, total acquisition time = 4 min.

The reconstruction can be formulated as:
$$\hat{I} = \arg\min_I \|k - MF^{-1}CI\|_2 + \lambda R \quad [Eq. 6]$$
where $k$ is the multi-channel under-sampled k-space measurement, $I$ is the 4D complex-valued multi-echo GRE images to be reconstructed, $M$ is the CAIPI undersampling mask, $F^{-1}$ is an inverse Fourier transform operator, and $C$ is the coil sensitivity map.

Conventional image reconstruction methods based on standard gradient descent least-square approaches require defining Eq.[6] as a linear function with a transpose operator so that the optimisation problem can be solved iteratively via least-square solvers (e.g., *lsqr*) or preconditioned conjugate gradients solvers. Tikhonov regularisation can also be added as the regulariser $R$. For comparison, we reconstructed the data both with and without Tikhonov regularisation using MATLAB's *lsqr* function. To reduce the memory usage, we first applied a 1D fast Fourier transform to the readout direction of $k$ and the reconstruction was performed on a slice-by-slice basis.

Correspondence to Kwok-Shing Chan

We reconstructed the data using the *askadam.m* solver in *GACELLE* using a similar expression:

$$\hat{I} = \arg\min_I \|k - MF^{-1}CI\|_{L1|L2} + \lambda \, |\nabla_{2D} I|_{L1} \quad [Eq. 7]$$

Reconstructions were performed using both L1-norm and L2-norm as the loss functions, respectively. Additionally, we incorporated 2D-TV regularisation for the L1-norm in the optimisation process. The reconstructed multi-echo data were subsequently used to compute the $R_2^*$ map using a closed-form approach (Gil et al., 2016) and QSM using the following pipeline: ROMEO (Dymerska et al., 2021) for echo combination, VSHARP (Li et al., 2011) for background field removal and LPCNN (Lai et al., 2020) for dipole inversion, as implemented in SEPIA (Chan and Marques, 2021) to illustrate the effect of incorporating spatial regularisation in image reconstruction for downstream analysis.

# 3. Results

## 3.1. Speed test

On CPU, the average runtimes were 0.022s per sample for NLLS and 4.6s per sample for $MH_{CPU}$ on AxCaliberSMT (Figure 4a), and 0.091s per sample for NLLS and 28.3s per sample for $MH_{CPU}$ on NEXI (Figure 4b). For the GPU-based stochastic gradient descent solver (*askadam.m*), the runtime was initially slower than CPU at the smallest sample size (acceleration factor $R_{AxCaliberSMT}$ = 0.14 and $R_{NEXI}$ = 0.84 for $10^2$ samples). However, scalability improved substantially with larger workloads, reaching $R_{AxCaliberSMT}$ = 54 and $R_{NEXI}$ = 451 for $2.5 \times 10^5$ samples.

For the stochastic inference solver *mcmc.m*, GPU processing was consistently faster than CPU across all sample sizes. The ensemble sampler achieved the highest acceleration for AxCaliberSMT, peaking at $R_{AxCaliberSMT}$ = 2027 for $2.5 \times 10^5$ samples. Its performance plateaued earlier than with $MH_{GPU}$, suggesting full GPU resource (cores and memory) utilisation at smaller workloads. Although $MH_{GPU}$ was slower than the ensemble sampler for small datasets, it reached a comparable acceleration to the ensemble sampler with $R_{AxCaliberSMT}$ = 1942 for $2.5 \times 10^5$ samples. For NEXI, the highest acceleration was achieved by $MH_{GPU}$ ($R_{AxCaliberSMT}$ = 14,380), with the ensemble sampler close behind ($R_{AxCaliberSMT}$ = 12,212) at $2.5 \times 10^5$ samples.

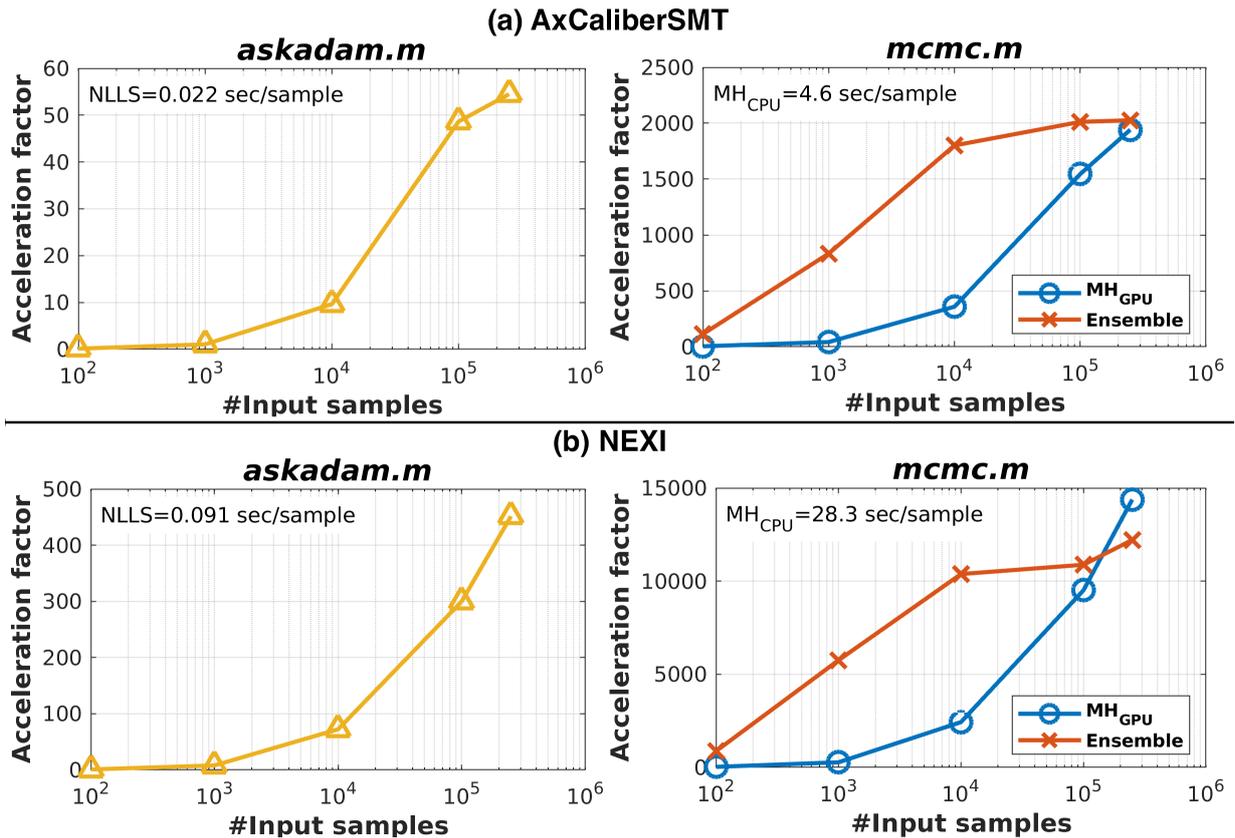

Figure 4: Acceleration factors achieved with *GACELLE* solvers relative to CPU implementations on (a) AxCaliberSMT and (b) NEXI. (Left) Runtime acceleration of the stochastic gradient descent solver *askadam.m* as a function of the number of samples processed in parallel on the GPU. (Right) Runtime acceleration of the stochastic inference solver *mcmc.m*, comparing the ensemble sampler and $MH_{GPU}$ across different sample sizes.

### 3.2. NEXI volumetric fitting with *askadam.m*

Volumetric fitting required an average of 18.5 CPU-hours using NLLS, whereas GPU-based optimisation with *askadam.m* reduced processing time to 1.3-2.5 minutes per subject, depending on the applied regularisation strategy. Bland-Altman analysis indicated a weak bias of exchange time (0.7 ms) between NLLS estimation and *askadam.m* without any regularisation, with additional small differences (0.2-0.4 ms) when spatial TV was applied (Figure 5a). The limits of agreement (LoA) between NLLS and *askadam.m* were relatively wider (1.96×SD = 15.4 ms), with NLLS tending to estimate longer exchange time in some ROIs.

CoV of exchange time was highest for NLLS across the cortex, whereas CoV decreased substantially when spatial TV regularisation was applied, particularly for 3D TV (Figure 5b). More cortical regions also showed a lower IQR of exchange time estimates within the ROI when TV was applied (Figure 5c). Notably, elevated exchange time contrast observed in motor, somatosensory, visual and cingulate cortices was reduced after applying 3D spatial regularisation (black arrows, Figure 5b). Note that with 3D TV regularisation, the group-averaged exchange time was also reduced in the posterior cingulate/retrosplenial cortices, from which the NLLS's CoV was relatively low (orange Arrows, Figure 5b), demonstrating the risks of over-smoothing that can be introduced by spatial regularisation.

### 3.3. NEXI vertex fitting with *askadam.m*

Vertex-based fitting across both hemispheres required approximately 8.2 CPU-hours using NLLS, whereas GPU-accelerated optimisation with *askadam.m* reduced this to 8.8-10.6 minutes per subject. All methods showed similar group-averaged exchange time contrast maps, but CoV reduction was clearly visible when surface TV regularisation or prior-based regularisation was applied (Figure 5d). ROIs that exhibited long estimated exchange time aligned with cortical regions known to be highly myelinated from high spatial resolution "myelin maps" (Glasser and Essen, 2011), e.g., primary visual, motor, auditory, and retrosplenial cortices. Importantly, contrasts observed in these ROIs were less affected by the application of regularisation when compared to volumetric fits interpolated to the same cortical depth. Incorporating a prior distribution for lower SNR data yielded the smallest mean absolute difference (MAD = 2.22ms) relative to high-SNR reference data, compared with MAD >3.8 ms for other fitting methods.

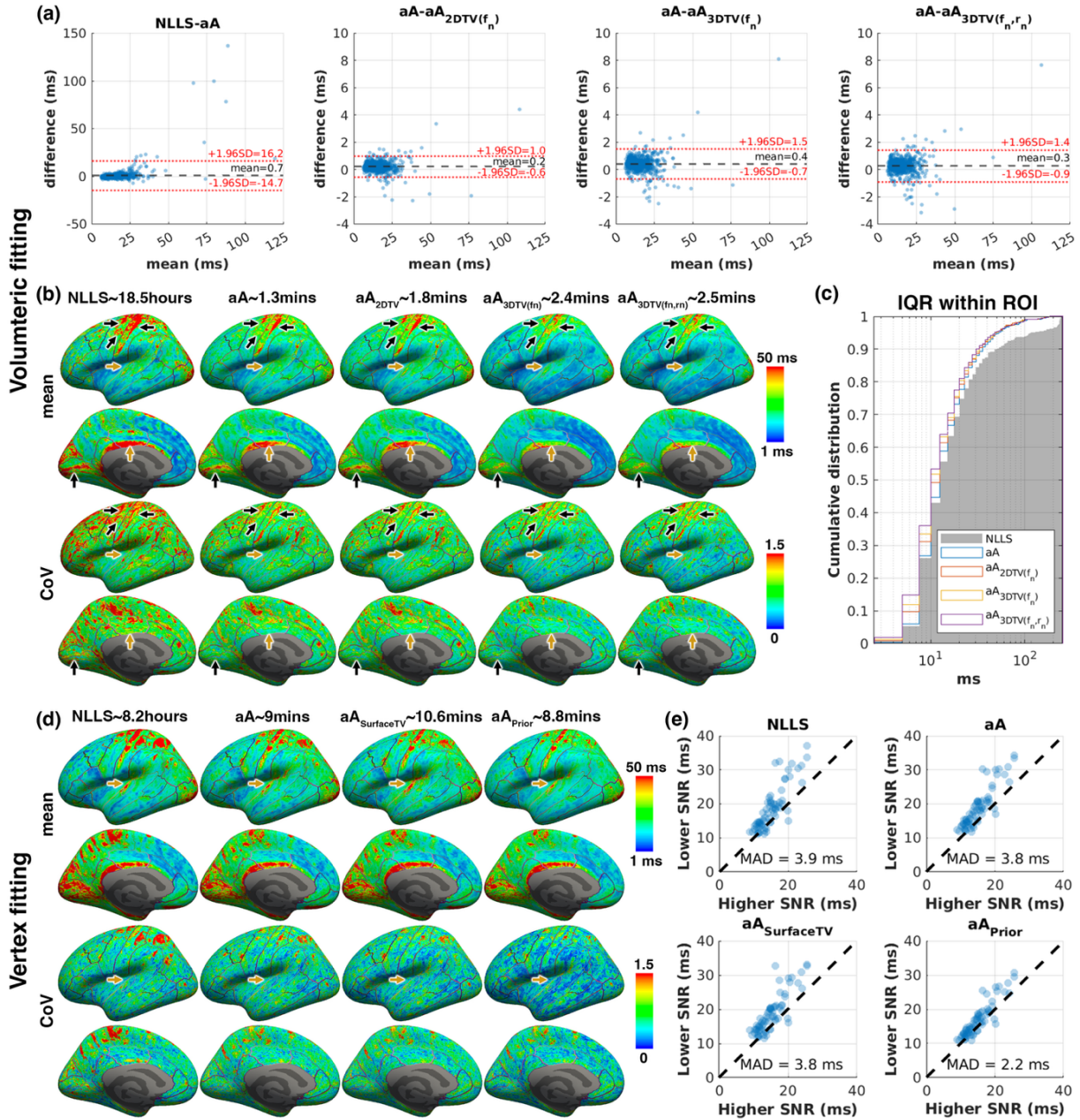

Figure 5: Group-level analysis of NEXI exchange time estimation by performing volumetric fitting (a-c) and vertex-based fitting (d,e). (a) Bland-Altman plots comparing volumetric fitting with various solvers. (b) Group-averaged exchange time (top) and CoV (bottom) projected onto the left cortical surface. Black arrows: regions with both high mean and CoV; orange arrows: retrosplenial and auditory cortices, where mean values were reduced by TV regularisation, despite low CoV. (c) Cumulative distribution of ROI-wise IQR from various solvers. (d) Group-averaged vertex-wise exchange time and CoV from surface-based fitting. (e) Scatter plots comparing the exchange time estimate from higher-SNR (using ordinary *askadam.m*) and lower SNR datasets across solvers. Each point represents a cortical ROI; prior-based regularisation yields the smallest group-wise mean absolute difference (MAD). NLLS: Non-linear least square; aA: *askadam.m*; 2DTV/3DTV: total variation regularisation in 2D/3D; SurfaceTV: surface-based total variation regularisation; Prior: prior-based regularisation.

## 3.4. AxCaliberSMT estimation with *mcmc.m*

The average processing time per subject was approximately 330 hours on a single CPU core ($MH_{CPU}$), compared to 12 minutes with *GACELLE*'s $MH_{GPU}$ implementation and 10 minutes using the ensemble sampler. All MCMC implementations produced comparable AxCaliberSMT parameter estimates (Figure 6). $MH_{GPU}$ showed no bias relative to $MH_{CPU}$ (Figure 6a and b). Estimation biases and LoA for $MH_{GPU}$ were similar to those obtained when $MH_{CPU}$ was repeated with independent proposal schemes (Figure 6b and c).

Axon diameter estimates from $MH_{CPU}$ and $MH_{GPU}$ were also comparable in group analysis (Figure 6d and e). However, LoA were wider for the MH sampler compared to the ensemble sampler, primarily driven by outlier ROIs near the grey/white matter boundary (black arrows, Figure 6d and f).

All MCMC implementations exhibited similar IQR distributions across white matter ROIs, indicating comparable uncertainty levels in group-level analysis (Figure 6g).

## 3.5. MCR-MWI estimation with intractable forward model

*GACELLE*'s *askadam.m* implementation of MCR-MWI took approximately 20 minutes to process a whole-brain dataset, whereas the original CPU implementation was about 250 CPU-hours. Highly myelinated tracts, such as the optic radiation (OR), were clearly visible in the MWF maps estimated by *askadam.m* but not in NLLS fits, matching the higher-SNR appearance baseline maps observed when diffusion-derived priors were incorporated in MCR-MWI (orange arrows, Figure 7a). Good agreement can be observed between *askadam.m* and NLLS in white matter tract ROIs (Pearson r = 0.92; Figure 7b), with minimal bias (0.26 %; Figure 7c) and narrow LoA ([-0.70,1.21] %; Figure 7c). Test-retest reproducibility also improved across all protocols, with correlations increasing from r = 0.73-0.74 for NLLS to r = 0.85-0.87 for *askadam.m*, accompanied by reduced dispersion across sessions (Figure 7d).

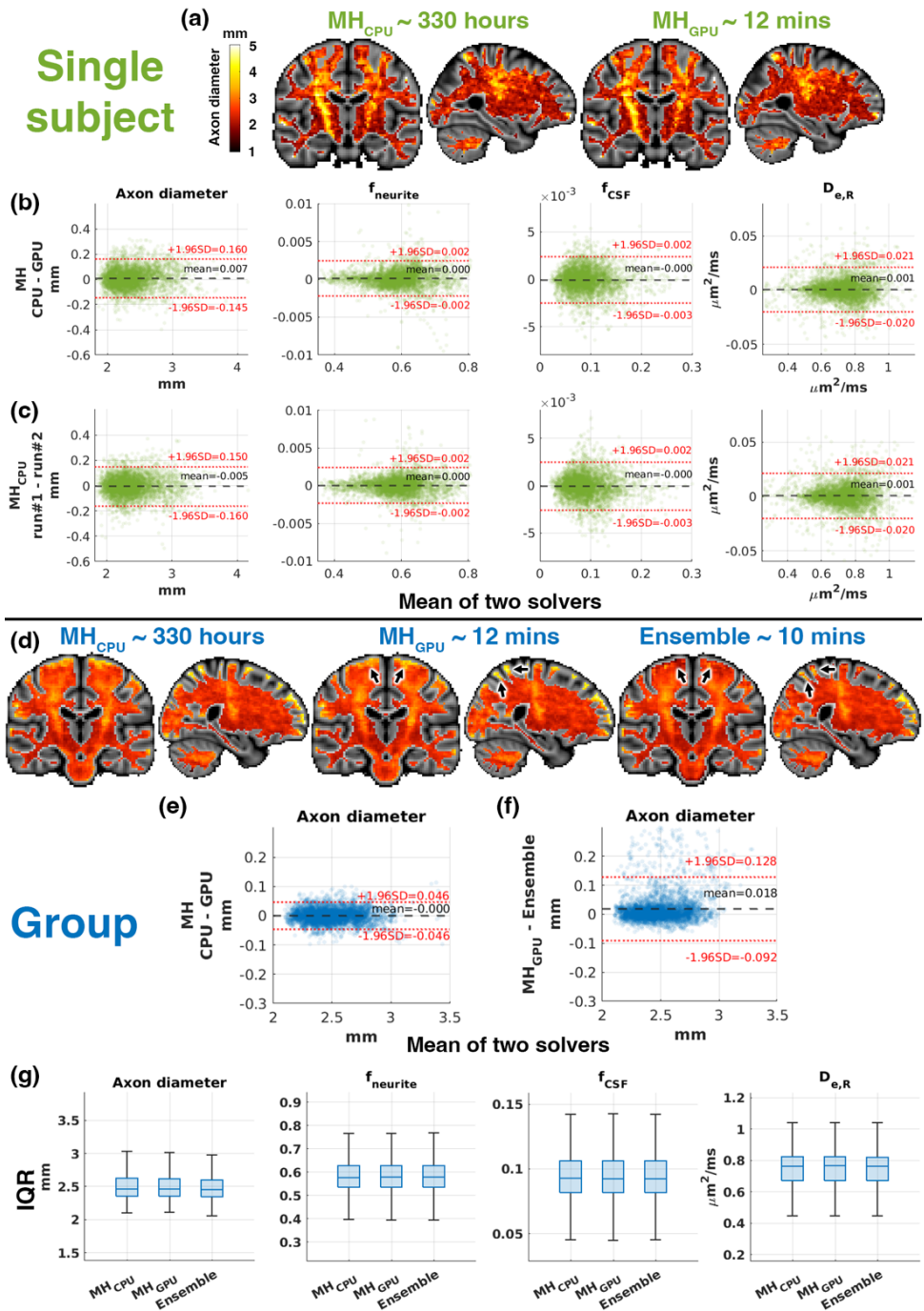

Figure 6: Comparison of AxCaliberSMT parameter estimation using *mcmc.m* in *GACELLE* across CPU and GPU implementation. (a) Example axon diameter maps from $MH_{CPU}$ and $MH_{GPU}$ in one subject. (b) Bland-Altman plots of all tissue parameters (axonal diameter, neurite signal fraction $f_{neurite}$, free water signal fraction $f_{CSF}$ and radial diffusivity of extracellular water $D_{e,R}$) between $MH_{CPU}$ and $MH_{GPU}$. (c) Bland-Altman plots between two independent proposal schemes on $MH_{CPU}$. (d) Group-averaged axon diameter maps across solvers. (e) Bland-Altman plot comparing group-averaged axon diameter between $MH_{CPU}$ and $MH_{GPU}$. (f) Bland-Altman plot comparing $MH_{GPU}$ and ensemble sampler, with the wider LoA associated with partial volume effect near tissue boundaries (black arrows). (g) Box plots of ROI-level IQR for all tissue parameters across solvers, showing comparable uncertainty distributions.

Correspondence to Kwok-Shing Chan

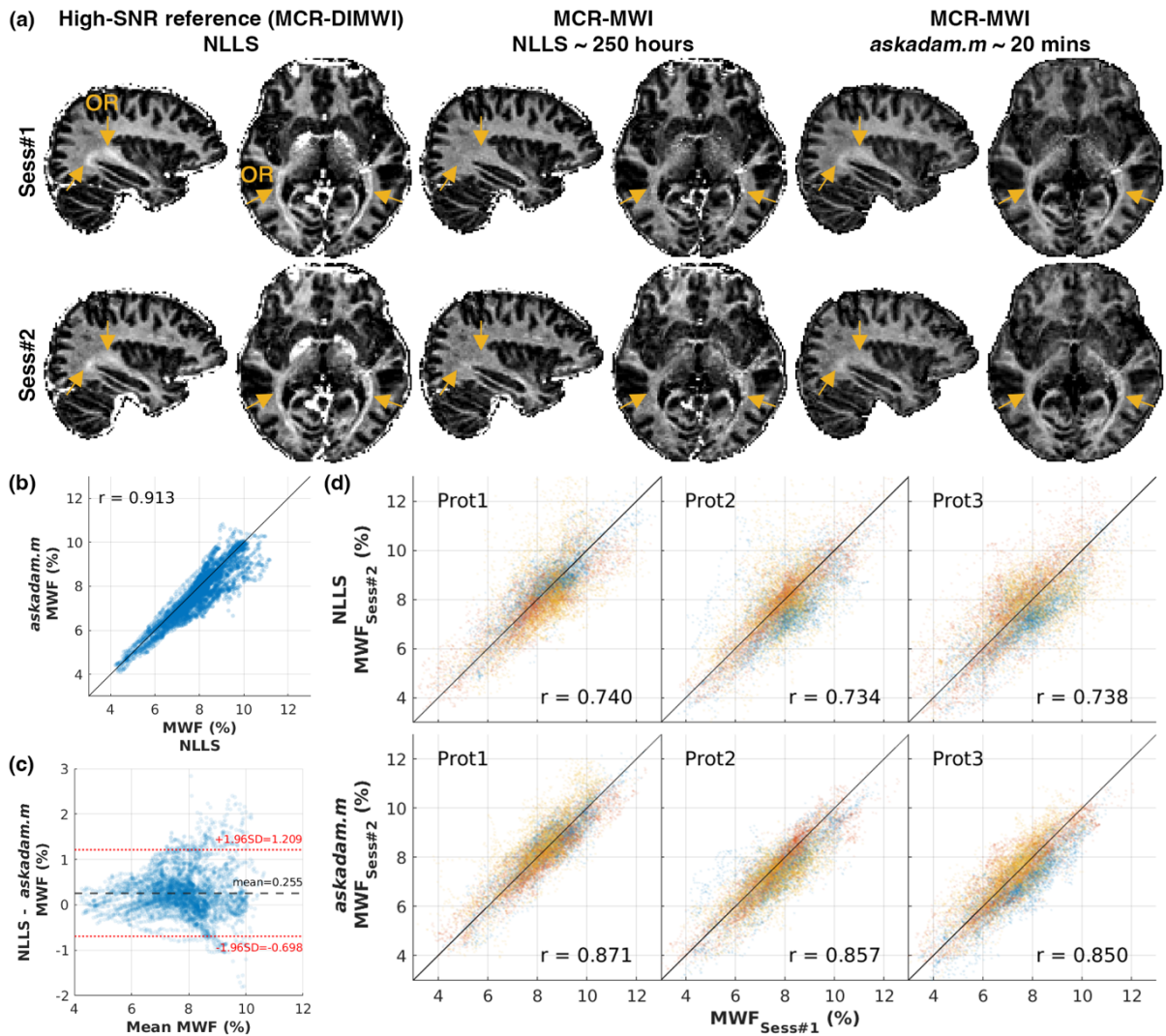

Figure 7: MWF estimation from NLLS and *askadam.m* in *GACELLE*. (a) High-SNR reference MWF map incorporating diffusion MRI microstructure information (left), voxelwise NLLS estimate (middle), and *askadam.m* estimate (right). The optic radiation (OR) is more clearly defined in the *askadam.m* map, consistent with the higher-SNR reference (orange arrows). (b) Scatter plot and (c) Bland-Altman plot comparing MCR-MWI MWF estimates between NLLS and *askadam.m*. (d) Test-retest scatter plots for three imaging protocols showing higher reproducibility for *askadam.m* compared to NLLS. The three colours correspond to three different subjects.

### 3.6. Using *askadam.m* solver for image reconstruction

The reconstructed images between conventional optimisation methods (based on *lsqr*) and *askadam.m* were comparable (Figure 8a). Incorporating spatial TV regularisation with *askadam.m* during reconstruction visibly reduced noise in the magnitude images, as expected (Figure 8a). Absolute difference maps showed a pattern consistent with a typical g-factor noise map, and the magnitude is comparable across echoes. Downstream analysis demonstrated that $R_2^*$ and QSM maps from TV-regularised reconstructions were smoother, with structural boundaries more clearly defined, particularly in $R_2^*$ (orange arrows, Figure 8b). Absolute difference maps for $R_2^*$ were consistent with a random noise pattern, while QSM differences contained more structural variation due to the non-local nature of susceptibility reconstruction (Figure 8c).

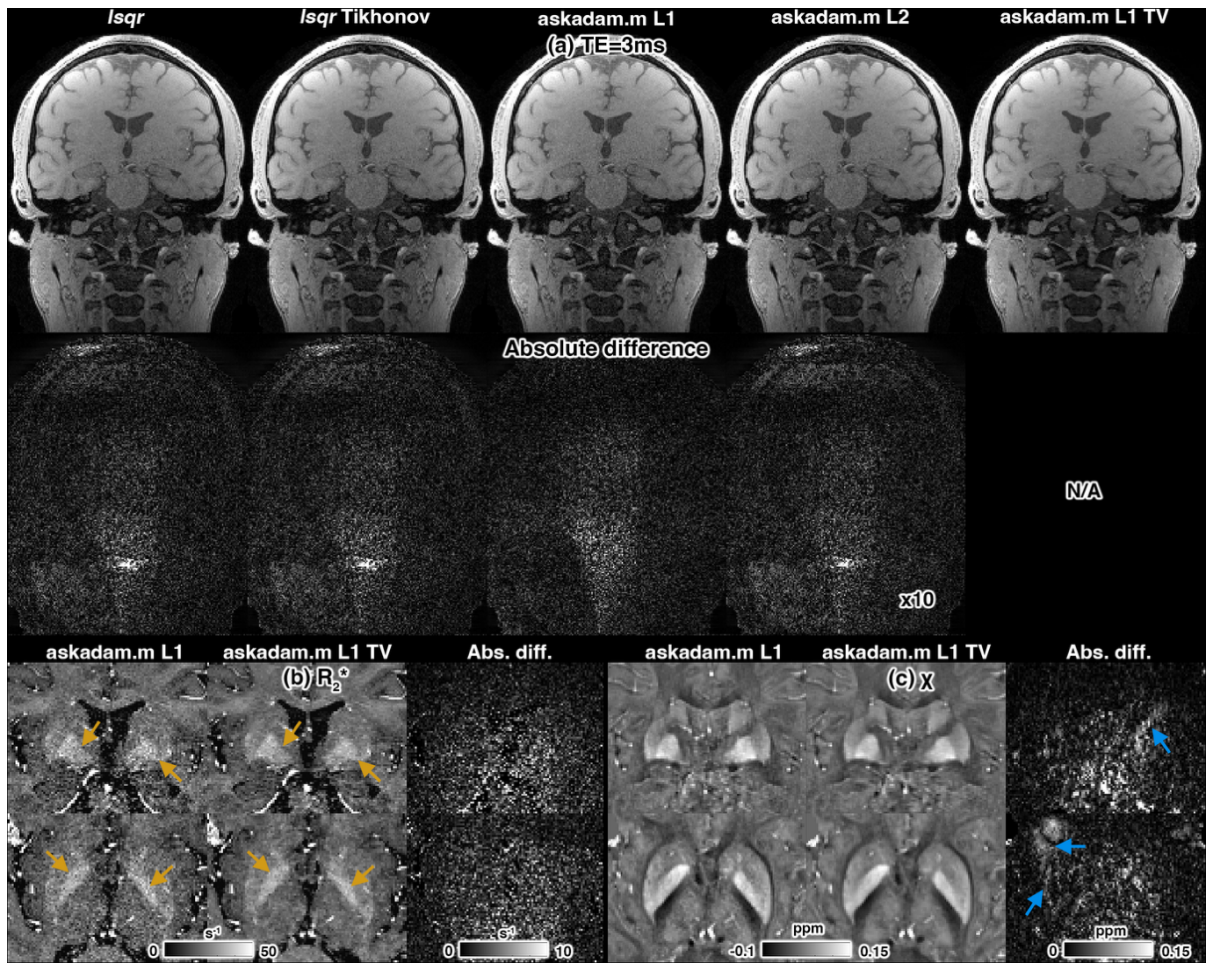

Figure 8: Multi-echo GRE reconstruction with various reconstruction methods, including linearised LSQR with and without L2-Tikhonov regularisation, *askadam.m* with L1 and L2 losses, and L1 loss with 2D TV regularisation, respectively. (a) Magnitude of the first echo images. Reconstructions with TV regularisation show reduced noise. Absolute difference maps (magnified 10 times) confirm residuals consistent with g-factor noise and are displayed on the same intensity range. (b) Zoomed $R_2$* maps and (c) magnetic susceptibility maps. TV-regularised reconstruction shows smoother $R_2$* maps and clearer tissue boundaries (orange arrows), with differences in susceptibility reflecting the non-local nature of QSM processing (blue arrows).

# 4. Discussion

In this work, we introduced *GACELLE*, a GPU-accelerated framework for high-throughput qMRI parameter estimation, implemented natively in MATLAB. *GACELLE* integrates both a stochastic gradient descent solver and a stochastic inference sampler within a single flexible interface, enabling large-scale parameter estimation, the incorporation of spatial regularisation, and general optimisation tasks such as image reconstruction. Compared to CPU-based implementations, *GACELLE* achieved acceleration factors up to 451 times for the stochastic gradient descent solver and 14,380 times for the stochastic inference sampler estimation on our tested models, while maintaining quantitative agreement with conventional solvers across multiple models, including NEXI, AxCaliberSMT, and MCR-MWI. This combination of speed, flexibility, and ease of integration makes *GACELLE* a practical tool for accelerating both routine and advanced qMRI workflows, potentially enabling the quick implementation of new and more complex biophysical models that can be readily applicable to higher resolution studies and larger cohort analyses.

Fast and robust parameter estimation remains a critical barrier to the broader adoption of qMRI. Previous work has explored GPU-accelerated least squares fittings (often in CUDA or Python) as well as deep learning–based surrogates (Harms et al., 2017; Harms and Roebroeck, 2018; Hernandez-Fernandez et al., 2019; Jung et al., 2021; Lee et al., 2019; Nedjati-Gilani et al., 2017). Compared to deep learning approaches, *GACELLE* achieves speed performance in the same order of magnitude while retaining the interpretability and generalisability of conventional model-based fitting. Our speed test results indicated that *GACELLE* runtime performance was maximum when handling a large dataset, given the allowance of the GPU resource (Figure 4). Unlike artificial neural networks, which often require re-optimisation for each acquisition protocol or scanner vendor (Jung et al., 2021; Lee et al., 2019), *GACELLE*'s solvers are less sensitive to sequence variations. By remaining model-based, *GACELLE* also preserves parameter interpretability and reproducibility across studies, which is essential for longitudinal or multisite applications. *GACELLE* provides a complementary option to existing tools for researchers who work in MATLAB and want to accelerate their workflow, emphasising accessibility for those without extensive GPU programming expertise. Memory usage and backend data management were optimised to enable efficient large-scale processing. Users can flexibly select loss functions, optimisers, and forward models without implementing GPU code themselves, leaving only the definition of the forward model in their hands. In our experience, we found that using L1-norm as the loss function for *askadam.m* generally provides a more stable convergence with fewer outliers compared to using L2-norm. While using L2-norm can produce similar performance as using L1-norm, as shown in the image reconstruction demo, some algorithm parameters, such as *initialLearnRate* and *tol*, need to be adapted to accommodate the change of loss computation. In addition, the stochastic gradient descent solver supports spatial regularisation, allowing users to leverage neighbourhood information to reduce noise while retaining biologically meaningful variance. Nonetheless, as shown in Figure 5b, when incorporating 3D TV on multiple parameters, the regularisation effect not only smoothed out the ROIs with high variance (motor and somatosensory cortices) but also the ROI with low variance (posterior cingulate gyrus/retrosplenial cortices). Therefore, careful tuning of the regularisation parameter remains essential to avoid over-smoothing relevant features.

Beyond qMRI parameter estimation, *GACELLE* can be applied to a wide range of optimisation problems in MRI, including complex-valued image reconstruction (Figure 8). Many reconstruction tasks require handling large, multi-dimensional datasets, integrating coil sensitivity encoding, non-Cartesian trajectories, or non-local operations such as dipole convolutions in QSM. *GACELLE*'s GPU-accelerated backend supports these operations directly within the optimisation loop, enabling efficient iterative reconstructions. In our demonstration, spatial TV regularisation was incorporated into a highly accelerated multi-echo GRE reconstruction. This approach visibly reduced noise in $R_2^*$ and QSM maps, sharpened tissue boundaries, and improved downstream quantitative analysis. The ability to integrate arbitrary forward models with custom regularisation makes *GACELLE* well-suited for other model-based reconstructions, such as joint parameter mapping and reconstruction (Sbrizzi et al., 2018; Zhao et al., 2015), dynamic imaging with temporal constraints (Otazo et al., 2015; Tamir et al., 2017), or physics-informed deep priors (Aggarwal et al., 2019; Mani et al., 2021). Although not shown in this manuscript, the ability to perform joint fitting of the whole imaging matrix, together with spatial regularisation, enables a straightforward joint fitting of QSM and $R_2^*$ to complex-valued GRE data (Marques et al., 2023), similar to total field inversion (Liu et al., 2017). By decoupling the optimisation framework from any single application domain, *GACELLE* provides a flexible platform that can be adapted to diverse MRI reconstruction challenges without requiring low-level GPU programming.

*GACELLE* currently requires MATLAB (with a paid license) and supports only NVIDIA GPUs, which may limit accessibility. Future work could include a precompiled standalone version compatible with MATLAB Runtime, so that paid licenses will not be required for its usage. Although memory optimisation reduces overhead, very large datasets (e.g., high-resolution and/or high-dimensional data) or models with many parameters may still approach GPU memory limits. Future versions could address this by implementing memory-efficient batching or multi-GPU execution that is currently supported by MATLAB. In addition, not all MATLAB built-in functions support *dlarray* objects for *askadam.m*, which may require users to implement alternative operations for some advanced image processing tasks. The global optimisation formalism used for *askdam.m* can also yield minor voxelwise differences from traditional nonlinear least squares, though our results suggest these differences are negligible for in vivo data. Additionally, since all voxels are included in a single loss metric, a few poorly fitted outliers may terminate the optimisation prematurely due to apparent convergence; careful selection of masks and fitting boundaries is therefore needed to mitigate this risk. Finally, the advantage of *GACELLE* is in its computational efficiency and usability. However, if the signal model (with the data available) is ill-conditioned, estimation using *askadam.m* or *mcmc.m* will likely inherit a similar problem, such as the degeneracy problem in parameter landscape (Jelescu et al., 2015). Recent work has shown promising results by incorporating a trained neural network to overcome this issue (Jallais and Palombo, 2024). *GACELLE* takes a different approach to handle this issue with the stochastic gradient descent solver, from which users can design an appropriate regularisation method to alleviate the ill-conditioned problem.

Correspondence to Kwok-Shing Chan

Future directions will focus on integrating prior knowledge, such as hierarchical Bayesian models (Orton et al., 2014), into the MCMC framework to improve estimation robustness. For stochastic gradient descent solvers, combining deep learning–based priors (e.g., convolutional neural network denoisers) with model-based fitting could provide a powerful balance between learning driven regularisation and data fidelity as done in variational networks with unrolled optimisation (Hammernik et al., 2018). Incorporating *GACELLE* with a physics-informed neural network (PINN) (Raissi et al., 2019) may also extend *GACELLE* from a model-based optimisation tool to tackle other biophysical problems, such as solving ordinary differential equations in arterial spin labelling (Galazis et al., 2025). Enhanced GPU memory management tools will also support scalability across diverse GPU devices, including potential for on-scanner implementation. Together, these developments could position *GACELLE* as a platform for hybrid model-based and learning-based estimation, extending its applicability from research pipelines to prospective clinical deployment.

Correspondence to Kwok-Shing Chan

## 5. Conclusions

We developed *GACELLE*, a MATLAB native GPU-accelerated framework for high-throughput qMRI parameter estimation, integrating a stochastic gradient descent solver and a stochastic inference sampler within a flexible interface. Across representative models, including NEXI, AxCaliberSMT, and MCR-MWI, *GACELLE* achieved acceleration factors up to 14,380 relative to CPU-based tools while maintaining quantitative accuracy with conventional solvers. The framework supports spatial regularisation, handles complex-valued optimisation problems, and extends naturally to general MRI reconstruction tasks. By combining speed, flexibility, and accessibility for MATLAB users, *GACELLE* provides a practical solution for accelerating model-based quantitative MRI analyses and enabling larger-scale, higher-resolution, and more reproducible studies.

Correspondence to Kwok-Shing Chan

# Acknowledgements

KC is supported by NWO/ZonMw under the Rubicon award number 04520232330012. Hansol Lee was supported by the Bio&Medical Technology Development Program of the National Research Foundation funded by the Korean government (MSIT) (No. RS-2024-00411768). This study was support by the Office of the Director (OD) of the National Institutes of Health (NIH) in partnership with the National Institute of Dental and Craniofacial Research (NIDCR) under award number DP5OD031854, the OD of the NIH under award number S10OD032184, the National Institute of Neurological Disorders and Stroke (NINDS) of NIH under award numbers R01NS118187, the National Institute of Biomedical Imaging and Bioengineering (NIBIB) under award numbers P41EB015896, P41EB030006, and U01EB026996, the National Center for Research Resources (NCRR) of NIH under award numbers S10RR023401 and S10RR019307, and the National Institute on Aging under award number R21AG085795. BB is supported by NIH under award numbers R01EB032378, R01EB028797, UH3EB034875, R21AG082377 and R01EB034757. JM is supported by NWO Open Technology Program 2024 - Quantitative Imaging that (white) Matters: Taking Biophysical Modelling to Clinical Application - https://doi.org/10.61686/CJTDV86349.

Correspondence to Kwok-Shing Chan

Correspondence to Kwok-Shing Chan

Correspondence to Kwok-Shing Chan

Correspondence to Kwok-Shing Chan

Correspondence to Kwok-Shing Chan

Correspondence to Kwok-Shing Chan

# Data and Code Availability

The data and code that support the findings of this study will be made available upon reasonable request. The GPU processing tool for the model fitting and image reconstruction is available at https://github.com/kschan0214/gacelle.

Correspondence to Kwok-Shing Chan

# Author Contributions

**Kwok-Shing Chan**: Conceptualisation, Methodology, Software, Validation, Formal analysis, Investigation, Writing – Original Draft, Visualisation, Funding acquisition

**Hansol Lee**: Software, Investigation, Writing – Review & Editing

**Yixin Ma**: Investigation, Writing – Review & Editing

**Berkin Bilgic**: Conceptualisation, Methodology, Writing – Review & Editing

**Susie Y. Huang**: Methodology, Writing – Review & Editing, Supervision, Funding acquisition

**Hong-Hsi Lee**: Conceptualisation, Methodology, Software, Writing – Original Draft, Supervision, Funding acquisition

**José P. Marques**: Conceptualisation, Methodology, Software, Writing– Original Draft, Supervision, Funding acquisition

Correspondence to Kwok-Shing Chan

# Declaration of Competing Interests

The authors have no competing interests related to the findings of this work.

Correspondence to Kwok-Shing Chan

# Supplementary Materials

Table S1: Acquisition parameters of the in vivo data used for Demo #1-3.

| Demo | #1 NEXI | #2 AxCaliberSMT | | #3 MCR-MWI | | |
|---|---|---|---|---|---|---|
| Scanner | Connectome 2.0 | | Scanner | Prisma | | |
| | DWI | | | Variable-flip-angle-multi-echo GRE | | |
| Diffusion time, $D$ (ms) | [13,21,30] | [13,30] | Protocol | #1 | #2 | #3 |
| #diff. grad. Dir per shells | 32 (lower SNR) or 64 (high SNR) | 32 or 64 | Flip angle (°) | [5,10,20,50,70] | | |
| $b$ (ms/$\mu$m$^2$) | [2.3,3.5,4.8,6.5, 11.3*,17.3**] *$D\geq$21ms; **$D$=30ms | $D_{13}$=[0.05,0.35,0.8,1.5, 2.4,3.45,4.75,6]; $D_{30}$=[0.2,0.95,2.3,4.25, 6.75,9.85,13.5,17.8] | TR/TE$_1$/$\Delta$TE(ms)/ #TE | 38/2.2/3.07/ 12 | 50/2.2/3.07/ 15 | 55/2.68/3.95/ 13 |
| TR/TE (ms) | 3600/54 | | Data type | Complex-valued | | |
| SMS/R$_{GRAPPA}$/PF | 2/2/0.75 | | R$_{CAIPI}$ | 5 | | |
| Res. (mm) | 2 (isotropic) | | Res. (mm) | 1.5 (isotropic) | | |
| TA (min) | 40 (lower SNR) or 80 (high SNR) | 56 | TA (min) | 14 | 18 | 20 |

Correspondence to Kwok-Shing Chan

# A. Semi-supervised learning for fast multi-compartment relaxometry myelin water imaging (MCR-MWI)

## Methods

In multi-compartment relaxometry for myelin water imaging (MCR-MWI) (Chan and Marques, 2020), variable flip angle data introduce $T_1$ weighting that helps separate myelin water (MW) from intra-/extra-cellular water (IEW), as these compartments exhibit different $T_1$ relaxation. Imperfect gradient and RF spoiling preclude a simple analytical steady state, therefore, the compartmental $T_1$-weighted steady-state signal is modelled with the extended phase graph with exchange (EPG-X) formalism (Malik et al., 2017) rather than a Bloch–McConnell closed-form solution. Direct use of EPG-X within voxelwise nonlinear least squares (NLLS) is computationally demanding; without parallelisation, a single whole-brain dataset at 1.5-mm isotropic resolution requires approximately 250 CPU-hours.

To enable fast, parallel evaluation of the EPG-X steady-state signal within *GACELLE*, we trained an artificial neural network (ANN) surrogate to approximate the compartmental steady-state signal. The trained network is embedded in the MCR-MWI forward model and provides fast evaluations during optimisation across large volumes without subject-specific training, maintaining applicability across a range of acquisition settings.

## ANN for EPG-X steady-state $T_1$ weighted signals

### Architecture and inputs

We used a multi-layer perceptron (MLP) with 7 hidden layers [20, 30, 40, 45, 60, 65, 75] and leaky RELU activations (scale factor = 0.01), as illustrated in Figure S1. The network accepts 6 inputs: myelin volume fraction $f_M$, intra-/extra-axonal $T_1$ ($T_{1,IEW}$),

intra-/extra-axonal T$_2$ ($T_{2,IEW}$), exchange rate ($k_{IEWM}$), flip angle ($\alpha$), and repetition time (TR). Feature transformations were applied to these inputs to produce 11 normalised features. The model contains 15782 learnable parameters. Using the same input format and architecture, two networks were trained: one predicts the differences between the IEW and MW steady-state magnitude signal derived from EPG-X and Bloch-McConnell solution; the other predicts the phase of the IEW steady-state signal, each to match the corresponding EPG-X simulation.

## Training data

We generated $2\times10^6$ random parameter sets $\theta$ with the following parameter ranges: $f_M \in [0,0.72]$, $T_{1,IEW} \in [500,6000]$ ms, $T_{2,IEW} \in [25,4000]$ ms, $k_{IEWM} \in [0,10]$ s$^{-1}$, TR $\in$ [25, 85] ms, $\alpha \in [1,90]°$. Myelin T$_1$ and T$_2$ were fixed to 234 ms and 15 ms, respectively. The RF spoiling phase was set to 50°. For each $\theta$, steady-state signals were simulated for $\alpha$ from 1° to 90° (step size = 2.3°).

## Training procedures and loss

Networks were trained for 100 epochs using the Adam optimiser. The total loss was the sum of three L1 terms:

$$total\ loss = L1_\theta + L1_{\theta,\alpha\in[1-90]°} + \lambda L1_{dS/d\alpha}\ [Eq.S1]$$

with

$$L1_\theta = \|\Delta S_{EPG-X}(\theta) - S_{ANN}(\theta)\|_1\ [Eq.S2a]$$

$$L1_{\theta,\alpha\in[1-90]°} = \left\|\Delta S_{EPG-X_{\alpha\in[1-90]°}}(\theta,\alpha) - S_{ANN_{\alpha\in[1-90]°}}(\theta,\alpha)\right\|_1, \text{aggregated over } \alpha \in [1-90]°\ [Eq.S2b]$$

$$L1_{dS/d\alpha} = \left\|\frac{d\,\Delta S_{EPG-X_{\alpha\in[1-90]°}}(\theta)}{d\alpha} - \frac{d\,S_{ANN_{\alpha\in[1-90]°}}(\theta)}{d\alpha}\right\|_1, \text{aggregated over } \alpha \in [1-90]°\ [Eq.S2c]$$

where $\Delta S_{EPG-X}$ is the signal difference between EPG-X steady-state and Bloch-McConnell solution. The first term enforces fidelity at each $\theta$; the second matches the full flip angle response; the third term aligns the first derivative with respect to $\alpha$, encouraging smooth, physically plausible steady-state curves. The hyperparameter $\lambda$ was initialised at 100 and reduced adaptively across training epochs as $\lambda_i = 100/(1 + 0.1(i-1))$ where $i$ is the epoch number. The network parameters were trained with a hybrid strategy of increasing batch size and reducing learning rate as a function of the training iterations to ensure convergence.

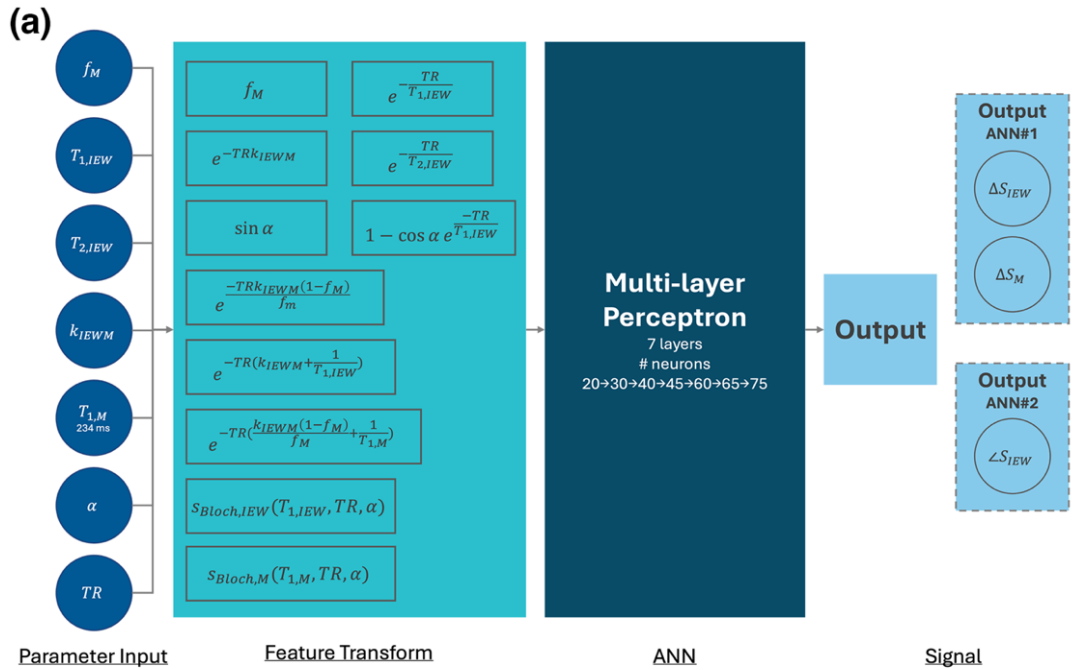

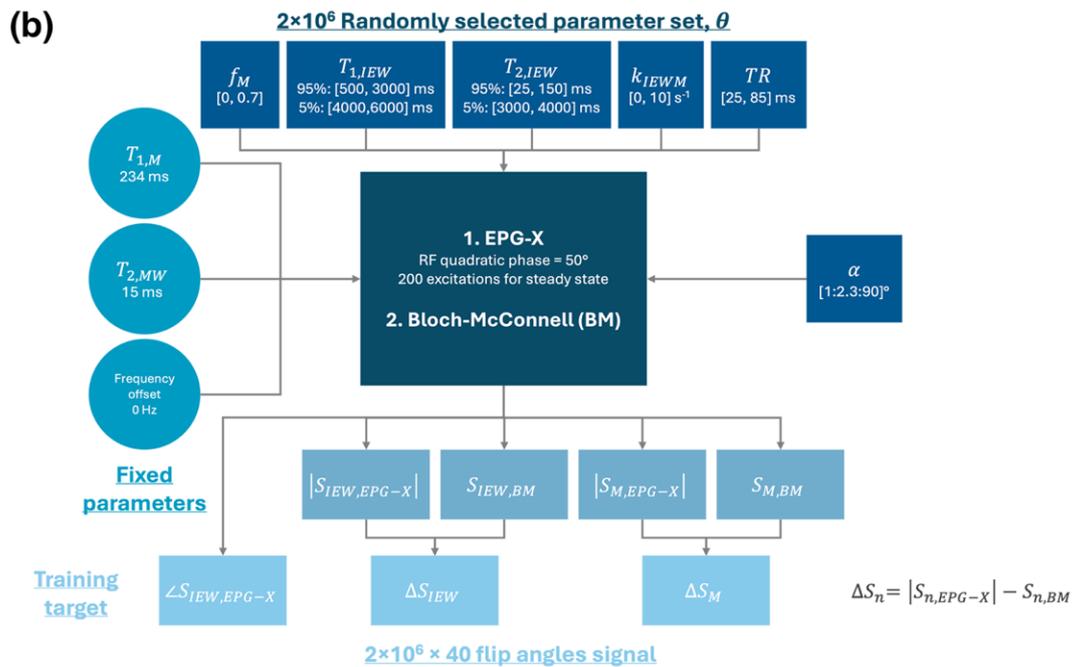

Figure S1: a) The EPG-X steady-state ANN is preceded by a feature transformation step to derive 11 input features that are both normalised to [0,1] and are related to different terms of the Bloch-McConnell

Correspondence to Kwok-Shing Chan

equations. The network core comprises 7 hidden fully-connected layers and a leaky RELU (scale factor=0.01) function. The network parameters are trained with a hybrid strategy of increasing batch size and reducing learning rate as a function of the training iterations to ensure convergence. Note that the first ANN was trained to learn the difference between the magnitude EPG-X and Bloch-McConnell signals, while the second ANN was trained to learn the phase of the IEW signal. b) Illustration of the parameters and their ranges to generate training and validation data.

## Integration into *GACELLE*

After training, the ANN replaces the EPG-X computation block within the MCR-MWI forward model used by askadam.m. This preserves model fidelity through the training targets while enabling efficient GPU-accelerated optimisation over entire volumes. No subject-specific retraining is required; users specify acquisition parameters in the forward model, and *GACELLE* performs vectorised evaluation, gradient-based updates, and optional spatial regularisation during fitting.

## Validation

To assess the performance of the ANNs, we generated steady-state signals across flip angles from 1° to 90° using three approaches: EPG-X, Bloch-McConnell (BM) equation, and the ANNs, applied to a range of tissue and acquisition parameter sets (Figure S2). In this two-pool model, the BM signal of IEW usually exhibits stronger bias relative to EPG-X, primarily due to prolonged $T_2$ decay. Across all tested scenarios, the ANN output showed substantially reduced bias compared to BM, indicating that the ANN provides an effective correction for RF spoiling effects. For the phase of the IEW signal, the ANN reproduced the overall trend of the EPG-X simulation, with deviation increasing at higher flip angle and exchange rate; nonetheless, these differences remained within 10% of the true phase values. For the myelin signal, where the short $T_2$ yields smaller discrepancies between EPG-X and BM, residuals were generally low. Even in this case, ANN correction further reduced the residual bias, demonstrating that the ANN can capture subtle deviations overlooked by the BM formulation.

Correspondence to Kwok-Shing Chan

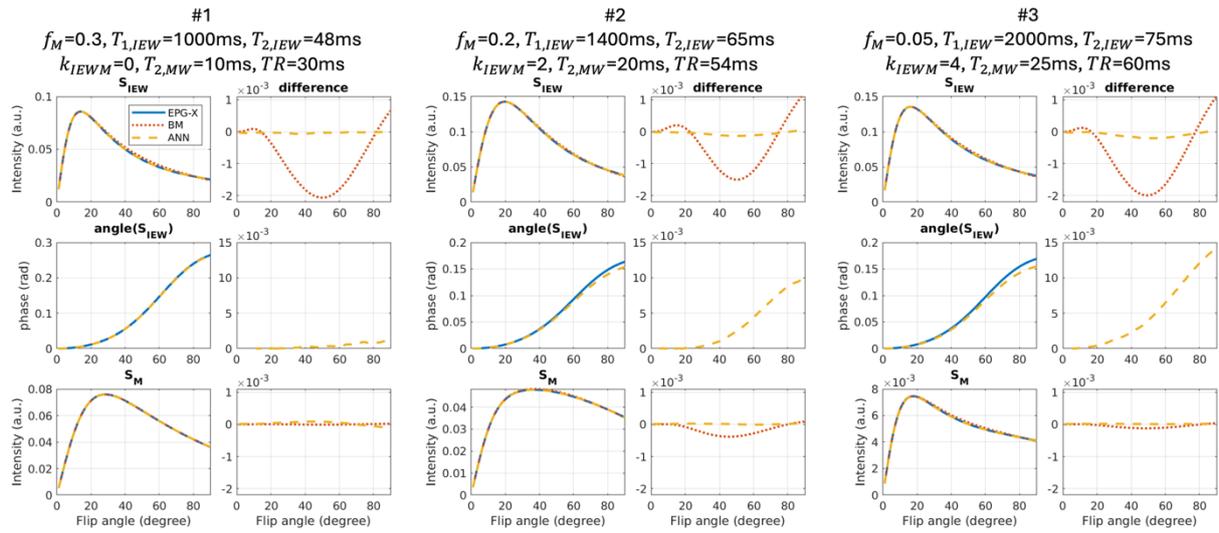

Figure S2: Steady-state signals derived from EPG-X, Bloch-McConnell equation, and ANN in three representative parameter sets. Comparisons highlight the reduced bias of ANN relative to BM, particularly for the IEW compartment.

Correspondence to Kwok-Shing Chan